\documentclass[10pt]{article}
\usepackage{graphicx}
\usepackage{amsmath}
\usepackage{amssymb}
\usepackage{caption2}
\begin{document}
\begin{center}
{\large {\bf \sc{ The decay widths  of the  $Z_{cs}(3985/4000)$ based on rigorous quark-hadron duality
  }}} \\[2mm]
Zhi-Gang  Wang \footnote{E-mail: zgwang@aliyun.com.  }   \\
 Department of Physics, North China Electric Power University, Baoding 071003, P. R. China
\end{center}

\begin{abstract}
In this work, we explore the hadronic coupling constants $G_{ZJ/\psi K}$, $G_{Z\eta_c K^*}$, $G_{ZD^* \bar{D}_s}$ of the exotic states $Z_{cs}(3985/4000)$ both in the pictures of the tetraquark states and molecular states with the tentative assignments $J^{PC}=1^{+-}$ based on the rigorous quark-hadron duality.   Then we  obtain the total widths $\Gamma_{Z_{cs}}^T =15.31\pm 2.06\,\rm{MeV}$ and $\Gamma_{Z_{cs}}^M=83.51\pm21.09\,\rm{MeV}$,
which are consistent with the experimental data $13.8^{+8.1}_{-5.2}\pm4.9\,\rm{MeV}$ from the BESIII collaboration  and
$131 \pm 15 \pm 26\,\rm{MeV}$ from the LHCb collaboration, respectively, and  support assigning the $Z_{cs}(3985)$ and $Z_{cs}(4000)$ to be the hidden-charm tetraquark state and molecular state with the $J^{PC}=1^{+-}$, respectively.
\end{abstract}



\section{Introduction}
In 2013, the BESIII and Belle collaborations  explored   the process  $e^+e^- \to \pi^+\pi^-J/\psi$, and observed a structure $Z_c^\pm(3900)$ in the $\pi^\pm J/\psi$ mass spectrum \cite{BES3900,Belle3900}.
Also in 2013, the BESIII collaboration investigated   the process $e^+e^- \to \pi D \bar{D}^*$, and observed a structure $Z_c^\pm(3885)$  in the $(D \bar{D}^*)^{\pm}$
 mass spectrum  \cite{BES-3885}. The  $Z_c(3900/3885)$ have the spin-parity  $J^P=1^+$ \cite{BES-3885,JP-BES-Zc3900}.

In 2020, the BESIII collaboration observed a structure $Z_{cs}^-(3985)$ in the $K^{+}$ recoil-mass spectrum  in the processes of the $e^+e^-\to K^+ (D_s^- D^{*0} + D^{*-}_s D^0)$ \cite{BES3985}.   The Breit-Wigner  mass and width  are  $3985.2^{+2.1}_{-2.0}\pm1.7\,\rm{MeV}$   and $13.8^{+8.1}_{-5.2}\pm4.9\,\rm{MeV}$, respectively  \cite{BES3985}.

In 2021, the LHCb collaboration observed two new  exotic states $Z_{cs}^+(4000)$ and $Z_{cs}^+(4220)$ in the $J/\psi K^+$  mass spectrum  in the process  $B^+ \to J/\psi \phi K^+$ \cite{LHCb-Zcs4000}.  The most significant state, $Z_{cs}^+(4000)$, has the Breit-Wigner  mass and width $4003 \pm 6 {}^{+4}_{-14}\,\rm{MeV}$ and $131 \pm 15 \pm 26\,\rm{MeV}$, respectively,  and the spin-parity $J^P =1^+$ \cite{LHCb-Zcs4000}.

The $Z_c(3900/3885)$ and $Z_{cs}(3985/4000)$ have analogous decay modes,
\begin{eqnarray}
Z^{\pm}_c(3900) &\to & J/\psi \pi^\pm \, , \nonumber \\
Z^{+}_{cs}(4000) &\to & J/\psi K^+ \, ,
\end{eqnarray}
\begin{eqnarray}
Z_c^{\pm}(3885)&\to & (D\bar{D}^*)^\pm \, , \nonumber \\
Z_{cs}^{-}(3985) &\to & D_s^- D^{*0}\, , \,  D^{*-}_s D^0 \, ,
\end{eqnarray}
which lead to the possible outcome that  they should have analogous quark structures. Although the Particle Data Group takes it for granted that the $Z_c(3900)$ and $Z_c(3885)$ are  the same particle according to the  analogous masses and widths \cite{PDG}, however, it is difficult to explain the ratio
 \cite{BES-3885},
 \begin{eqnarray}
 R_{exp}&=&\frac{\Gamma(Z_c(3885)\to D\bar{D}^*)}{\Gamma(Z_c(3900)\to J/\psi \pi)} =6.2 \pm 1.1 \pm 2.7 \, ,
 \end{eqnarray}
in a satisfactory way. On the other hand,  the widths of the  $Z^-_{cs}(3985)$ and $Z_{cs}^+(4000)$
  are inconsistent with each other. The $Z_c(3900/3885)$ and $Z_{cs}(3985/4000)$ may be four distinguished  particles indeed.

The exotic  states $Z_{cs}(3985/4000)$ lie above  the $D_s^- D^{*0}$ and $D^{*-}_s D^0$ thresholds $3975.2\,\rm{MeV}$ and $3977.0\,\rm{MeV}$,  respectively,
 if they are molecular states indeed, we should introduce the coupled-channel effects to account for the mass gaps and decay widths \cite{Zcs-molecule-1,Zcs-molecule-2,Zcs-molecule-3,Zcs-molecule-4,Zcs-molecule-5}, or just re-scattering effects  \cite{Chen:2013wca,Zcs-Rescatt-1}. It is more natural to reproduce their masses in the diquark-antidiquark type  tetraquark pictures than in the color-singlet-color-singlet type tetraquark pictures,  if we choose the theoretical framework of  potential quark models \cite{Ebert:2008kb, Ferretti:2020ewe}.

In the QCD sum rules, we usually choose the color antitriplet-triplet type and singlet-singlet type four-quark currents to interpolate the hidden-charm exotic states, and can reproduce the masses of the $Z_{cs}(3985/4000)$ both in the pictures of the tetraquark states \cite{Zcs-CFQiao,Zcs-tetra-quark,WZG-Zcs3985-tetra,Zcs3985-Azizi,Ozdem-Zcs-tetra-mole} and molecular states \cite{Zcs-CFQiao,Ozdem-Zcs-tetra-mole,Lee:2008uy,Dias:2013qga,Zcs-molecule-6,WZG-Zcs3985-mole}. There maybe exist two $Z_{cs}$ states, an  antitriplet-triplet type tetraquark state and singlet-singlet type molecular state, or only one $Z_{cs}$ state, which has both the antitriplet-triplet type and singlet-singlet type components. We have to explore the two-body strong decays to diagnose their nature.

In Ref.\cite{WZG-ZJX-Zc-Decay}, we assign the $Z^\pm_c(3900)$  to be the  diquark-antidiquark type tetraquark state with the quantum numbers $J^{PC}=1^{+-}$,  calculate the hadronic coupling constants $G_{ZJ/\psi\pi}$, $G_{Z\eta_c\rho}$, $G_{ZD \bar{D}^{*}}$ with  the three-point QCD sum rules based on the rigorous  duality for the first time,   then obtain  the partial decay widths to diagnose the nature of the $Z_c^\pm(3900)$. Thereafter, the rigorous  duality has been successfully applied to study the strong decays of the exotic states $X(4140)$, $X(4274)$, $X(4660)$, $Z_c(4600)$ and $P_c(4312)$ \cite{WZG-Y4660-Decay,WZG-X4140-decay,WZG-X4274-decay,WZG-Z4600-decay,WZG-Pc4312-decay-mole,WZG-Pc4312-decay-tetra}.

In this work, we extend our previous work on the masses of the hidden-charm tetraquark states and molecular states with the strangeness  \cite{WZG-Zcs3985-tetra,WZG-Zcs3985-mole}, to explore the two-body strong decays of the   $Z_{cs}(3985/4000)$ with the possible assignments $J^{PC}=1^{+-}$, so as to diagnose their nature.

The article is arranged as follows:  we acquire  the QCD sum rules for the hadronic coupling constants $G_{ZJ/\psi K}$, $G_{Z\eta_cK^*}$, $G_{ZD^* \bar{D}_s}$ in section 2; in Sect.3, we present the numerical results and discussions; and Sect.4 is reserved for our
conclusion.

\section{The hadronic coupling constants of the $Z_{cs}(3985/4000)$ }

We  investigate the two-body strong decays $Z_{cs}^-(3985/4000)\to J/\psi K^-$, $\eta_cK^{*-}$, $D^{*0} \bar{D}_s^{-}$, $D^0 \bar{D}_s^{*-}$ using  the three-point correlation functions $\Pi_{\mu\nu}^{1}(p,q)$, $\Pi_{\mu\nu}^{2}(p,q)$ and $\Pi_{\mu\nu}^{3}(p,q)$, respectively,
\begin{eqnarray}
\Pi_{\mu\nu}^{1}(p,q)&=&i^2\int d^4xd^4y \, e^{ipx}e^{iqy}\, \langle 0|T\left\{J_\mu^{J/\psi}(x)J_5^{K}(y)J^\dagger_{\nu}(0)\right\}|0\rangle\, ,  \nonumber \\
\Pi_{\mu\nu}^{2}(p,q)&=&i^2\int d^4xd^4y\, e^{ipx}e^{iqy}\, \langle 0|T\left\{J_5^{\eta_c}(x)J_\mu^{K^*}(y)J^\dagger_{\nu}(0)\right\}|0\rangle \, , \nonumber \\
\Pi_{\mu\nu}^{3}(p,q)&=&i^2\int d^4xd^4y \, e^{ipx}e^{iqy}\, \langle 0|T\left\{J_\mu^{D^*}(x)J_5^{D_s}(y)J^\dagger_{\nu}(0)\right\}|0\rangle \, ,
\end{eqnarray}
where the currents
\begin{eqnarray}
J_\mu^{J/\psi}(x)&=&\bar{c}(x)\gamma_\mu c(x) \, ,\nonumber \\
J_5^{K}(y)&=&\bar{u}(y)i\gamma_5 s(y) \, , \nonumber \\
J_5^{\eta_c}(x)&=&\bar{c}(x)i\gamma_5 c(x) \, ,\nonumber \\
J_\mu^{K^*}(y)&=&\bar{u}(y)\gamma_\mu s(y) \, , \nonumber \\
J_\mu^{D^*}(x)&=&\bar{u}(x)\gamma_\mu c(x) \, ,\nonumber \\
J_5^{D_s}(y)&=&\bar{c}(y)i\gamma_5 s(y) \, , \nonumber \\
J^\dagger_\nu(0)&=&J_\nu^{T\dagger}(0)\, , \, J_\nu^{M\dagger}(0)\, ,
\end{eqnarray}
\begin{eqnarray}
J_\nu^{T\dagger}(0)&=&\frac{\varepsilon^{ijk}\varepsilon^{imn}}{\sqrt{2}}\Big\{c^{Tn}(0)C\gamma_\nu u^m(0)\, \bar{c}^k(0)\gamma_5 C \bar{s}^{Tj}(0)-c^{Tn}(0)C\gamma_5 u^m(0)\, \bar{c}^k(0)\gamma_\nu C \bar{s}^{Tj}(0) \Big\} \, , \nonumber\\
J_\nu^{M\dagger}(0)&=& \frac{1}{\sqrt{2}}\Big\{\bar{s}(0)\gamma_\nu c(0)\, \bar{c}(0)i\gamma_5 u(0)+\bar{s}(0)i\gamma_5 c(0)\, \bar{c}(0) \gamma_\nu u(0)\Big\}\, ,
\end{eqnarray}
interpolate the mesons $J/\psi$, $K$, $\eta_c$, $K^*$, $D^*$, $D_s$, $Z^T_{cs}$ and $Z^M_{cs}$, respectively, the superscripts $T$ and $M$ denote the tetraquark-type and  molecule-type four-quark currents/tetraquarks, respectively \cite{WZG-Zcs3985-tetra,WZG-Zcs3985-mole}.

We insert  a complete set of intermediate hadronic states with the same quantum numbers as the currents into the three-point correlation functions, and  isolate the ground state contributions,
\begin{eqnarray}\label{Hadron-CT-1}
\Pi_{\mu\nu}^{T,1}
(p,q)&=&\left\{ \frac{f_{K}M_{K}^2f_{J/\psi}M_{J/\psi}\lambda_{Z}^T G^T_{ZJ/\psi K}}{m_u+m_s} \frac{i}{(M_{Z}^2-p^{\prime2})(M_{J/\psi}^2-p^2)(M_{K}^2-q^2)} \right. \nonumber\\
&&+\frac{i}{(M_{Z}^2-p^{\prime2})(M_{J/\psi}^2-p^2)}\int_{s^0_K}^{\infty} du\frac{\rho_{Z K^\prime}(p^{\prime2},p^2,u)}{u-q^2}\nonumber\\
&&+\frac{i}{(M_{Z}^2-p^{\prime2})(M_{K}^2-q^2)}\int_{s^0_{J/\psi}}^{\infty} ds\frac{\rho_{Z \psi^\prime}(p^{\prime2},s,q^2)}{s-p^2}\nonumber\\
&&\left.+\frac{i}{(M_{J/\psi}^2-p^2)(M_{K}^2-q^2)}\int_{s^0_Z}^{\infty} ds^\prime \frac{\rho_{Z^\prime J/\psi}(s^\prime,p^2,q^2)+\rho_{Z^\prime K}(s^\prime,p^2,q^2)}{s^\prime-p^{\prime2}}+\cdots \right\} g_{\mu\nu}+\cdots  \nonumber\\
&=& i\,\Pi_T^1(p^{\prime2},p^2,q^2)\,g_{\mu\nu}+\cdots\, ,
\end{eqnarray}

\begin{eqnarray}
\Pi_{\mu\nu}^{M,1}(p,q)&=& \left\{\frac{f_{K}M_{K}^2f_{J/\psi}M_{J/\psi}\lambda_{Z}^M G^M_{ZJ/\psi K}}{m_u+m_s} \frac{1}{(M_{Z}^2-p^{\prime2})(M_{J/\psi}^2-p^2)(M_{K}^2-q^2)} +\cdots \right\}g_{\mu\nu}+\cdots  \nonumber\\
&=& \Pi^1_M(p^{\prime2},p^2,q^2)\,g_{\mu\nu}+\cdots\, ,
\end{eqnarray}

\begin{eqnarray}
\Pi_{\mu\nu}^{T,2}(p,q)&=& \left\{\frac{f_{\eta_c}M_{\eta_c}^2f_{K^*}M_{K^*}\lambda_{Z}^TG_{Z\eta_c K^*}^T}{2m_c} \frac{i}{(M_{Z}^2-p^{\prime2})(M_{\eta_c}^2-p^2)(M_{K^*}^2-q^2)}+\cdots \right\} g_{\mu\nu}+\cdots  \nonumber\\
&=&i\, \Pi_T^2(p^{\prime2},p^2,q^2)\,g_{\mu\nu}+\cdots\, ,
\end{eqnarray}

\begin{eqnarray}
\Pi_{\mu\nu}^{M,2}(p,q)&=&\left\{ \frac{f_{\eta_c}M_{\eta_c}^2f_{K^*}M_{K^*}\lambda_{Z}^MG_{Z\eta_c K^*}^M}{2m_c} \frac{1}{(M_{Z}^2-p^{\prime2})(M_{\eta_c}^2-p^2)(M_{K^*}^2-q^2)} +\cdots \right\}g_{\mu\nu}+\cdots  \nonumber\\
&=& \Pi_M^2(p^{\prime2},p^2,q^2)\,g_{\mu\nu}+\cdots\, ,
\end{eqnarray}

\begin{eqnarray}
\Pi_{\mu\nu}^{T,3}(p,q)&=&\left\{ \frac{f_{D_s}M_{D_s}^2f_{D^*}M_{D^*}\lambda_{Z}^TG^T_{Z D^*\bar{D}_s}}{m_c+m_s} \frac{i}{(M_{Z}^2-p^{\prime2})(M_{D^*}^2-p^2)(M_{D_s}^2-q^2)} +\cdots \right\}g_{\mu\nu}+\cdots  \nonumber\\
&=&i\, \Pi_T^3(p^{\prime2},p^2,q^2)\,g_{\mu\nu}+\cdots \, ,
\end{eqnarray}

\begin{eqnarray}\label{Hadron-CT-6}
\Pi_{\mu\nu}^{M,3}(p,q)&=&\left\{ \frac{f_{D_s}M_{D_s}^2f_{D^*}M_{D^*}\lambda_{Z}^M G^M_{Z D^*\bar{D}_s}}{m_c+m_s} \frac{1}{(M_{Z}^2-p^{\prime2})(M_{D^*}^2-p^2)(M_{D_s}^2-q^2)}+\cdots\right\} g_{\mu\nu}+\cdots  \nonumber\\
&=& \Pi_M^3(p^{\prime2},p^2,q^2)\,g_{\mu\nu}+\cdots \, ,
\end{eqnarray}
where $p^\prime=p+q$, the $f_{J/\psi}$, $f_{K}$, $f_{\eta_c}$, $f_{K^*}$, $f_{D^*}$, $f_{D_s}$, $\lambda_Z^T$ and $\lambda_{Z}^M$  are the decay constants of the mesons  $J/\psi$, $K$, $\eta_c$, $K^*$, $D^*$, $D_s$,  $Z_{cs}^T$ and $Z_{cs}^M$, respectively,
\begin{eqnarray}
\langle0|J_{\mu}^{J/\psi}(0)|J/\psi(p)\rangle&=&f_{J/\psi}M_{J/\psi}\,\xi_\mu \,\, , \nonumber \\
\langle0|J_{\mu}^{K^*}(0)|K^*(q)\rangle&=&f_{K^*}M_{K^*}\,\varepsilon_\mu \,\, , \nonumber \\
\langle0|J_{\mu}^{D^*}(0)|D^*(p)\rangle&=&f_{D^*}M_{D^*}\,\varsigma_\mu \,\, , \nonumber \\
\langle0|J_{5}^{K}(0)|K(q)\rangle&=&\frac{f_{K}M_{K}^2}{m_u+m_s} \,\, , \nonumber \\
\langle0|J_{5}^{\eta_c}(0)|\eta_c(p)\rangle&=&\frac{f_{\eta_c}M_{\eta_c}^2}{2m_c} \,\, , \nonumber \\
\langle0|J_{5}^{D_s}(0)|D_s(q)\rangle&=&\frac{f_{D_s}M_{D_s}^2}{m_c+m_s} \,\, , \nonumber \\
\langle 0|J^{T/M}_\nu(0)|Z_{cs}^{T/M}(p^\prime)\rangle&=&\lambda_{Z}^{T/M}\,\zeta_\nu\,\, ,
\end{eqnarray}
and the $\xi$, $\varepsilon$, $\varsigma$ and $\zeta$ are polarization vectors of the $J/\psi$,  $K^*$, $D^*$ and $Z_{cs}^{T/M}$, respectively, the hadronic coupling constants
 $G^T_{ZJ/\psi K}$, $G^M_{ZJ/\psi K}$, $G^T_{Z\eta_cK^*}$, $G^M_{Z\eta_cK^*}$, $G_{Z D^*\bar{D}_s}^T$ and $G_{Z D^*\bar{D}_s}^M$ are defined by
\begin{eqnarray}
\langle J/\psi(p)K(q)|Z^T_{cs}(p^{\prime})\rangle&=&-\xi^*(p)\cdot\zeta(p^{\prime})\, G^T_{ZJ/\psi K} \, , \nonumber\\
\langle\eta_c(p)K^*(q)|Z_{cs}^T(p^{\prime})\rangle&=&-\varepsilon^*(q)\cdot\zeta(p^{\prime})\, G^T_{Z\eta_cK^*}  \, ,  \nonumber\\
\langle D^*(p)D_s(q)|Z^T_{cs}(p^{\prime})\rangle&=&-\varsigma^*(p)\cdot\zeta(p^{\prime})\, G^T_{Z D^*\bar{D}_s} \, ,   \nonumber\\
\langle J/\psi(p)K(q)|Z^M_{cs}(p^{\prime})\rangle&=&i\xi^*(p)\cdot\zeta(p^{\prime})\, G^M_{ZJ/\psi K} \, , \nonumber\\
\langle\eta_c(p)K^*(q)|Z_{cs}^M(p^{\prime})\rangle&=&i\varepsilon^*(q)\cdot\zeta(p^{\prime})\, G^M_{Z\eta_cK^*}  \, ,  \nonumber\\
\langle D^*(p)D_s(q)|Z^M_{cs}(p^{\prime})\rangle&=&i\varsigma^*(p)\cdot\zeta(p^{\prime})\, G^M_{Z D^*\bar{D}_s} \, .
\end{eqnarray}
The hadronic spectral densities $\rho_{Z K^\prime}(p^{\prime2},p^2,u)$, $\rho_{Z \psi^\prime}(p^{\prime2},s,q^2)$, $\rho_{Z^\prime J/\psi}(s^\prime,p^2,q^2)$, $\rho_{Z^\prime K}(s^\prime,p^2,q^2)$ in the component $\Pi_T^1(p^{\prime2},p^2,q^2)$ stand for the transitions between the ground states and continuum states (including the first radial excited states). In fact, such hadronic spectral densities also exist in the big $ \{  \}$ in the components $\Pi_T^2(p^{\prime2},p^2,q^2)$, $\Pi_T^3(p^{\prime2},p^2,q^2)$, $\Pi_M^1(p^{\prime2},p^2,q^2)$,
$\Pi_M^2(p^{\prime2},p^2,q^2)$, $\Pi_M^3(p^{\prime2},p^2,q^2)$, we neglect them for simplicity.
In this work, we choose the tensor structure  $g_{\mu\nu}$  to explore  the hadronic  coupling constants $G^{T/M}_{ZJ/\psi K}$, $G^{T/M}_{Z\eta_cK^*}$ and $G^{T/M}_{Z D^* \bar{D}_s}$, and neglect other tensor structures for simplicity.

We accomplish  the operator product expansion up to the vacuum condensates of dimension 5 and neglect the tiny gluon condensate contributions, just like in our previous works \cite{WZG-ZJX-Zc-Decay,WZG-Y4660-Decay,WZG-X4140-decay,WZG-X4274-decay,WZG-Z4600-decay}, then obtain the QCD spectral densities $\rho_{QCD}(p^{\prime2},s,u)$  through double dispersion relation, and write the correlation functions at the QCD side in the form,
\begin{eqnarray}
\Pi_{QCD}(p^{\prime2},p^2,q^2)&=& \int_{\Delta_s^2}^\infty ds \int_{\Delta_u^2}^\infty du \frac{\rho_{QCD}(p^{\prime2},s,u)}{(s-p^2)(u-q^2)}\, ,
\end{eqnarray}
where   the $\Delta_s^2$ and $\Delta_u^2$  are the thresholds.
While at the hadron side, we obtain the hadron  spectral densities $\rho_H(s^\prime,s,u)$ through triple  dispersion relation, and write the correlation functions  in the form,
\begin{eqnarray}
\Pi_{H}(p^{\prime2},p^2,q^2)&=&\int_{\Delta_s^{\prime2}}^\infty ds^{\prime} \int_{\Delta_s^2}^\infty ds \int_{\Delta_u^2}^\infty du \frac{\rho_{H}(s^\prime,s,u)}{(s^\prime-p^{\prime2})(s-p^2)(u-q^2)}\, ,
\end{eqnarray}
according to Eqs.\eqref{Hadron-CT-1}-\eqref{Hadron-CT-6}, where the $\Delta_{s}^{\prime2}$ are the thresholds.
We match the hadron side with the QCD side of the correlation functions bellow the continuum thresholds according to the dispersion relation at the QCD side  to warrant  rigorous quark-hadron  duality,
 \begin{eqnarray}
  \int_{\Delta_s^2}^{s_{0}}ds \int_{\Delta_u^2}^{u_0}du  \frac{\rho_{QCD}(p^{\prime2},s,u)}{(s-p^2)(u-q^2)}&=& \int_{\Delta_s^2}^{s_0}ds \int_{\Delta_u^2}^{u_0}du  \left[ \int_{\Delta_{s}^{\prime2}}^{\infty}ds^\prime  \frac{\rho_H(s^\prime,s,u)}{(s^\prime-p^{\prime2})(s-p^2)(u-q^2)} \right]\, ,
\end{eqnarray}
where  the $s_0$ and $u_0$ are the continuum thresholds,  we accomplish the integral over $ds^\prime$ firstly, and introduce some unknown parameters to parameterize the contributions involving the higher resonances and continuum states in the $s^\prime$ channel \cite{WZG-ZJX-Zc-Decay,WZG-Y4660-Decay}, we take account of the higher resonances and continuum states rather than neglecting them.
For example, according to Eq.\eqref{Hadron-CT-1}, we  write down the correlation function  at the hadron side explicitly,
 \begin{eqnarray}
  \int_{\Delta_s^2}^{s_{0}}ds \int_{\Delta_u^2}^{u_0}du  \frac{\rho_{QCD}(p^{\prime2},s,u)}{(s-p^2)(u-q^2)}&=& \frac{A_{ZJ/\psi K}}{(M_{Z}^2-p^{\prime2})(M_{J/\psi}^2-p^2)(M_{K}^2-q^2)}+\frac{C^T_{J/\psi K}}{(M_{J/\psi}^2-p^2)(M_{K}^2-q^2)}\, , \nonumber \\
\end{eqnarray}
where
\begin{eqnarray}
A_{ZJ/\psi K}&=& \frac{f_{K}M_{K}^2f_{J/\psi}M_{J/\psi}\lambda_{Z}^T G^T_{ZJ/\psi K}}{m_u+m_s}\, , \nonumber \\
C^T_{J/\psi K}&=&\int_{s^0_Z}^{\infty} ds^\prime \frac{\rho_{Z^\prime J/\psi}(s^\prime,p^2,q^2)+\rho_{Z^\prime K}(s^\prime,p^2,q^2)}{s^\prime-p^{\prime2}} \, ,
\end{eqnarray}
where the unknown parameter $C_{J/\psi K}^T$ parameterizes  the contributions involving the higher resonances and continuum states with the same quantum numbers as the $Z_{cs}$ (in the $s^\prime$ channel), the unknown parameters $C_{J/\psi K}^M$, $C_{\eta_c K^*}^T$, $C_{\eta_c K^*}^M$, $C_{D^* \bar{D}_s}^T$,
    and $C_{D^* \bar{D}_s}^M$ in Eqs.\eqref{JpsiK-Z-M}-\eqref{DsDv-Z-M} are implied in the same way.

We accomplish the integral over the variable $ds^\prime$ firstly, and do not need the continuum threshold parameters $s^{\prime }_0$, and the duality is   rigorous in the sense that we do not set $s_0=s^{\prime}_0=s^0_Z$ approximately \cite{Nielsen-Zc3900}. If we set $s_0=s_0^{\prime}=s^0_Z$ in the present case, then $\sqrt{s_0}> M_Z>\psi^\prime>\eta_c^\prime>M_{D_s^{*\prime}}>M_{D_s^\prime}$, the contaminations from the excited states are out of control. In fact, without accomplishing the integral over the variable $ds^\prime$ firstly, the representations at the QCD side and at the hadron side cannot match with each other, up to now, no one can approve that they match with each other without merging the $s$ and $s^\prime$ channels $s_0=s^{\prime}_0=s^0_Z$ at the hadron side by hand, to be serious, such approximations $s_0=s^{\prime}_0$ are wrong, as the $s$ and $s^\prime$ channels are quite different.

In the $s$ and $u$ channels, we deal with the conventional mesons, and choose the standard vector and pseudoscalar currents to interpolate them,   direct calculations indicate that we can reproduce their masses satisfactorily with the two-point QCD sum rules below the continuum thresholds $s_0$ and $u_0$, respectively, the quark-hadron  duality is  reliable.

We set $p^{\prime2}=p^2$ and $p^{\prime2}=4p^2$ in the correlation functions $\Pi^{1/2}_{T/M}(p^{\prime 2},p^2,q^2)$ and $\Pi^{3}_{T/M}(p^{\prime 2},p^2,q^2)$, respectively, and accomplish the double Borel transform in regard  to the variables $P^2=-p^2$ and $Q^2=-q^2$ respectively, then set the Borel parameters  $T_1^2=T_2^2=T^2$  to get the six   QCD sum rules,
\begin{eqnarray} \label{JpsiK-Z-T}
&&\frac{f_{K}M_{K}^2f_{J/\psi}M_{J/\psi}\lambda_{Z}^TG^T_{ZJ/\psi K}}{m_u+m_s}\frac{1}{M_{Z}^2-M_{J/\psi}^2} \left[ \exp\left(-\frac{M_{J/\psi}^2}{T^2} \right)-\exp\left(-\frac{M_{Z}^2}{T^2} \right)\right]\exp\left(-\frac{M_{K}^2}{T^2} \right) \nonumber\\
&&+C^T_{J/\psi K} \exp\left(-\frac{M_{J/\psi}^2+M_{K}^2}{T^2}  \right)=\frac{1}{64\sqrt{2}\pi^4}\int_{4m_c^2}^{s^0_{J/\psi}} ds \int_{0}^{u^0_{K}} du  \sqrt{1-\frac{4m_c^2}{s}} u\left(2s+4m_c^2-3m_s m_c\right)\nonumber\\
&&\exp\left(-\frac{s+u}{T^2} \right)-\frac{m_s\left[2\langle\bar{q}q\rangle+\langle\bar{s}s\rangle\right]}{24\sqrt{2}\pi^2}\int_{4m_c^2}^{s^0_{J/\psi}} ds \sqrt{1-\frac{4m_c^2}{s}}\left(s+2m_c^2\right)\exp\left(-\frac{s}{T^2}\right) \nonumber\\
&&-\frac{m_s\langle\bar{q}g_s\sigma Gq\rangle}{288\sqrt{2}\pi^2}\int_{4m_c^2}^{s^0_{J/\psi}} ds \frac{s+20m_c^2}{\sqrt{s\left(s-4m_c^2\right)}}  \exp\left(-\frac{s}{T^2}\right)\nonumber\\
&&-\frac{m_s\langle\bar{q}g_s\sigma Gq\rangle}{24\sqrt{2}\pi^2T^2}\int_{4m_c^2}^{s^0_{J/\psi}} ds \sqrt{1-\frac{4m_c^2}{s}}\left(s+2m_c^2\right) \exp\left(-\frac{s}{T^2}\right) \, ,
\end{eqnarray}

\begin{eqnarray} \label{JpsiK-Z-M}
&&\frac{f_{K}M_{K}^2f_{J/\psi}M_{J/\psi}\lambda_{Z}^MG^M_{ZJ/\psi K}}{m_u+m_s}\frac{1}{M_{Z}^2-M_{J/\psi}^2} \left[ \exp\left(-\frac{M_{J/\psi}^2}{T^2} \right)-\exp\left(-\frac{M_{Z}^2}{T^2} \right)\right]\exp\left(-\frac{M_{K}^2}{T^2} \right) \nonumber\\
&&+C^M_{J/\psi K} \exp\left(-\frac{M_{J/\psi}^2+M_{K}^2}{T^2}  \right)=\frac{1}{128\sqrt{2}\pi^4}\int_{4m_c^2}^{s^0_{J/\psi}} ds \int_{0}^{u^0_{K}} du  \sqrt{1-\frac{4m_c^2}{s}} u\left(2s+4m_c^2-3m_s m_c\right)\nonumber\\
&&\exp\left(-\frac{s+u}{T^2} \right)-\frac{m_s\left[2\langle\bar{q}q\rangle+\langle\bar{s}s\rangle\right]}{48\sqrt{2}\pi^2}\int_{4m_c^2}^{s^0_{J/\psi}} ds \sqrt{1-\frac{4m_c^2}{s}}\left(s+2m_c^2\right)\exp\left(-\frac{s}{T^2}\right)\nonumber\\
&&-\frac{m_s\langle\bar{q}g_s\sigma Gq\rangle}{288\sqrt{2}\pi^2}\int_{4m_c^2}^{s^0_{J/\psi}} ds \frac{s+20m_c^2}{\sqrt{s\left(s-4m_c^2\right)}} \exp\left(-\frac{s}{T^2}\right)\nonumber\\
&&-\frac{m_s\langle\bar{q}g_s\sigma Gq\rangle}{48\sqrt{2}\pi^2T^2}\int_{4m_c^2}^{s^0_{J/\psi}} ds \sqrt{1-\frac{4m_c^2}{s}}\left(s+2m_c^2\right)\exp\left(-\frac{s}{T^2}\right) \, ,
\end{eqnarray}

\begin{eqnarray}
&&\frac{f_{\eta_c}M_{\eta_c}^2f_{K^*}M_{K^*}\lambda_{Z}^TG^T_{Z\eta_c K^*}}{2m_c } \frac{1}{M_{Z}^2-M_{\eta_c}^2}\left[ \exp\left(-\frac{M_{\eta_c}^2}{T^2} \right)-\exp\left(-\frac{M_{Z}^2}{T^2} \right)\right]\exp\left(-\frac{M_{K^*}^2}{T^2} \right) \nonumber\\
&&+C^T_{\eta_c K^*} \exp\left(-\frac{M_{\eta_c}^2+M_{K^*}^2}{T^2}  \right)=-\frac{1}{64\sqrt{2}\pi^4}\int_{4m_c^2}^{s^0_{\eta_c}} ds \int_{0}^{u^0_{K^*}} du \sqrt{1-\frac{4m_c^2}{s}} u\left(2s-3m_sm_c \right)\nonumber\\
&&\exp\left(-\frac{s+u}{T^2}  \right)+\frac{m_s\langle\bar{q}q\rangle}{8\sqrt{2}\pi^2}\int_{4m_c^2}^{s^0_{\eta_c}} ds \sqrt{1-\frac{4m_c^2}{s}}\, s\, \exp\left(-\frac{s}{T^2}\right)\nonumber\\
&&+\frac{m_s\langle\bar{s}g_s\sigma Gs\rangle}{96\sqrt{2}\pi^2T^2}\int_{4m_c^2}^{s^0_{\eta_c}} ds \sqrt{1-\frac{4m_c^2}{s}}\, s\, \exp\left(-\frac{s}{T^2}\right)\nonumber\\
&&+\frac{m_s\langle\bar{q}g_s\sigma Gq\rangle}{72\sqrt{2}\pi^2}\int_{4m_c^2}^{s^0_{\eta_c}} ds \frac{s+2m_c^2}{\sqrt{s\left(s-4m_c^2\right)}}\exp\left(-\frac{s}{T^2}\right)\nonumber\\
&&-\frac{m_c\left[\langle\bar{q}g_s\sigma Gq\rangle+\langle\bar{s}g_s\sigma Gs\rangle\right]}{24\sqrt{2}\pi^2} \int_{4m_c^2}^{s^0_{\eta_c}} ds \sqrt{1-\frac{4m_c^2}{s}}\exp\left(-\frac{s}{T^2}\right) \, ,
\end{eqnarray}

\begin{eqnarray}
&&\frac{f_{\eta_c}M_{\eta_c}^2f_{K^*}M_{K^*}\lambda_{Z}^MG^M_{Z\eta_c K^*}}{2m_c } \frac{1}{M_{Z}^2-M_{\eta_c}^2}\left[ \exp\left(-\frac{M_{\eta_c}^2}{T^2} \right)-\exp\left(-\frac{M_{Z}^2}{T^2} \right)\right]\exp\left(-\frac{M_{K^*}^2}{T^2} \right) \nonumber\\
&&+C^M_{\eta_c K^*} \exp\left(-\frac{M_{\eta_c}^2+M_{K^*}^2}{T^2}  \right)=\frac{1}{128\sqrt{2}\pi^4}\int_{4m_c^2}^{s^0_{\eta_c}} ds \int_{0}^{u^0_{K^*}} du \sqrt{1-\frac{4m_c^2}{s}} u\left(2s-3m_sm_c \right)\nonumber\\
&&\exp\left(-\frac{s+u}{T^2}  \right)-\frac{m_s\langle\bar{q}q\rangle}{16\sqrt{2}\pi^2}\int_{4m_c^2}^{s^0_{\eta_c}} ds \sqrt{1-\frac{4m_c^2}{s}} s \exp\left(-\frac{s}{T^2}\right)\nonumber\\
&&-\frac{m_s\langle\bar{s}g_s\sigma Gs\rangle}{192\sqrt{2}\pi^2T^2}\int_{4m_c^2}^{s^0_{\eta_c}} ds \sqrt{1-\frac{4m_c^2}{s}}\,s\,\exp\left(-\frac{s}{T^2}\right)\nonumber\\
&&-\frac{m_s\langle\bar{q}g_s\sigma Gq\rangle}{72\sqrt{2}\pi^2} \int_{4m_c^2}^{s^0_{\eta_c}} ds \frac{s+2m_c^2}{\sqrt{s\left(s-4m_c^2\right)}} \exp\left(-\frac{s}{T^2}\right)\nonumber\\
&&+\frac{m_c\left[\langle\bar{q}g_s\sigma Gq\rangle+\langle\bar{s}g_s\sigma Gs\rangle\right]}{48\sqrt{2}\pi^2}\int_{4m_c^2}^{s^0_{\eta_c}} ds \sqrt{1-\frac{4m_c^2}{s}}\exp\left(-\frac{s}{T^2}\right) \, ,
\end{eqnarray}

\begin{eqnarray}
&&\frac{f_{D_s}M_{D_s}^2f_{D^*}M_{D^*}\lambda_{Z}^TG^T_{Z D^*\bar{D}_s}}{4(m_c+m_s) } \frac{1}{\widetilde{M}_{Z}^2-M_{D^*}^2}\left[ \exp\left(-\frac{M_{D^*}^2}{T^2} \right)-\exp\left(-\frac{\widetilde{M}_{Z}^2}{T^2} \right)\right]\exp\left(-\frac{M_{D_s}^2}{T^2} \right) \nonumber\\
&&+ C^T_{ D^*\bar{D}_s} \exp\left(-\frac{M_{D^*}^2+M_{D_s}^2}{T^2}  \right)=-\frac{m_c\langle\bar{s}g_s\sigma Gs\rangle}{96\sqrt{2}\pi^2}\int_{m_c^2}^{s^0_{D^*}} ds \left( \frac{9}{2}-\frac{10m_c^2}{s}+\frac{3m_c^4}{2s^2}\right)\exp\left(-\frac{s+m_c^2}{T^2}\right)\nonumber\\
&&+\frac{m_s\langle\bar{q}g_s\sigma Gq\rangle}{96\sqrt{2}\pi^2}\int_{m_c^2}^{u^0_{D_s}} du \left(2+\frac{3m_c^2}{u}\right) \frac{m_c^2}{u}\exp\left(-\frac{u+m_c^2}{T^2}\right)\nonumber\\
&&+\frac{m_c\langle\bar{q}g_s\sigma Gq\rangle}{96\sqrt{2}\pi^2}\int_{m_c^2}^{u^0_{D_s}} du \left( \frac{3}{2}+\frac{4m_c^2}{u}-\frac{3m_c^4}{2u^2}\right)\exp\left(-\frac{u+m_c^2}{T^2}\right)\nonumber\\
&&-\frac{m_s\langle\bar{q}g_s\sigma Gq\rangle}{192\sqrt{2}\pi^2}\int_{m_c^2}^{u^0_{D_s}} du \frac{1}{u-m_c^2}\left(u-9m_c^2-\frac{5m_c^4}{u}-\frac{3m_c^6}{u^2}\right)\exp\left(-\frac{u+m_c^2}{T^2}\right)\, ,
\end{eqnarray}

\begin{eqnarray} \label{DsDv-Z-M}
&&\frac{f_{D_s}M_{D_s}^2f_{D^*}M_{D^*}\lambda_{Z}^MG^M_{Z D^*\bar{D}_s}}{4(m_c+m_s) } \frac{1}{\widetilde{M}_{Z}^2-M_{D^*}^2}\left[ \exp\left(-\frac{M_{D^*}^2}{T^2} \right)-\exp\left(-\frac{\widetilde{M}_{Z}^2}{T^2} \right)\right]\exp\left(-\frac{M_{D_s}^2}{T^2} \right) \nonumber\\
&&+ C^M_{ D^*\bar{D}_s} \exp\left(-\frac{M_{D^*}^2+M_{D_s}^2}{T^2}  \right)=-\frac{3}{64\sqrt{2}\pi^4}\int_{m_c^2}^{s^0_{D^*}} ds \int_{m_c^2}^{u^0_{D_s}} du \left(1-\frac{m_c^2}{s}\right)\left(1-\frac{m_c^2}{u}\right)\nonumber\\
&&\left(2s-m_c^2-\frac{m_c^4}{s}\right)\left(u-m_c^2+2m_s m_c\right)\exp\left(-\frac{s+u}{T^2}\right)\nonumber\\
&&+\frac{m_c\langle\bar{s}s\rangle}{8\sqrt{2}\pi^2}\int_{m_c^2}^{s^0_{D^*}} ds \left(1-\frac{m_c^2}{s}\right)\left(2s-m_c^2-\frac{m_c^4}{s}\right)\exp\left(-\frac{s+m_c^2}{T^2}\right)\nonumber\\
&&+\frac{3m_c\langle\bar{q}q\rangle}{8\sqrt{2}\pi^2}\int_{m_c^2}^{u^0_{D_s}} du \left(1-\frac{m_c^2}{u}\right)\left(u-m_c^2+2m_s m_c\right)
\exp\left(-\frac{u+m_c^2}{T^2}\right)\nonumber\\
&&+\frac{m_s\langle\bar{s}s\rangle}{16\sqrt{2}\pi^2}\left(1+\frac{m_c^2}{T^2}\right)\int_{m_c^2}^{s^0_{D^*}} ds \left(1-\frac{m_c^2}{s}\right)\left(2s-m_c^2-\frac{m_c^4}{s}\right)\exp\left(-\frac{s+m_c^2}{T^2}\right)\nonumber\\
&&-\frac{3m_c^3\langle\bar{q}g_s\sigma Gq\rangle}{32\sqrt{2}\pi^2T^4}\int_{m_c^2}^{u^0_{D_s}} du \left(1-\frac{m_c^2}{u}\right)\left(u-m_c^2+2m_s m_c\right) \exp\left(-\frac{u+m_c^2}{T^2}\right)\nonumber\\
&&-\frac{m_sm_c^4\langle\bar{s}g_s\sigma Gs\rangle}{96\sqrt{2}\pi^2T^6}\int_{m_c^2}^{s^0_{D^*}} ds \left(1-\frac{m_c^2}{s}\right) \left(2s-m_c^2-\frac{m_c^4}{s}\right)\exp\left(-\frac{s+m_c^2}{T^2}\right) \nonumber\\
&&+\frac{m_c\langle\bar{s}g_s\sigma Gs\rangle}{16\sqrt{2}\pi^2T^2}\left(1-\frac{m_c^2}{2T^2} \right)\int_{m_c^2}^{s^0_{D^*}} ds\left(1-\frac{m_c^2}{s}\right) \left(2s-m_c^2-\frac{m_c^4}{s}\right)\exp\left(-\frac{s+m_c^2}{T^2}\right) \, ,
\end{eqnarray}	
where $\widetilde{M}_{Z}^2=\frac{M_{Z}^2}{4}$, the  $s^0_{J/\psi}$, $u^0_{K}$,  $s^0_{\eta_c}$, $u^0_{K^*}$, $s^0_{D^*}$ and $u^0_{D_s}$ are the continuum threshold parameters. There exist end-divergences at the endpoints $s=4m_c^2$ and $u=m_c^2$, we regularize the endpoint divergences with the simple replacements   $\frac{1}{\sqrt{s-4m_c^2}} \to \frac{1}{\sqrt{s-4m_c^2+4m_s^2}}$ and $\frac{1}{u-m_c^2} \to \frac{1}{u-m_c^2+4m_s^2}$   by adding a small squared $s$-quark mass $4m_s^2$ \cite{WZG-ZJX-Zc-Decay,WZG-Y4660-Decay,WZG-X4140-decay,WZG-X4274-decay,WZG-Z4600-decay}.

Now we take a short digression to illustrate the end-point divergences, we often encounter the typical
integral,
\begin{eqnarray}
I_{21}&=& \int d^4k_1
\frac{1}{(k_1^2-m_c^2)^2}\frac{1}{((p-k_1)^2-m_c^2)}\, ,
\end{eqnarray}
 and calculate it by using the Cutkosky's rules,
\begin{eqnarray}
I_{21}&=&\frac{\partial}{\partial A } \int d^4k_1\frac{1}{k_1^2-A}\frac{1}{(p-k_1)^2-m_c^2}\mid_{A\to m_c^2}\nonumber\\
&=&\frac{\partial}{\partial A } \frac{(-2\pi i)^2}{2\pi
i}\int_{(\sqrt{A}+m_c)^2}^{\infty}dt\frac{1}{t-p^2}\int d^4k_1k^4k_2\delta^4(k_1+k_2-p)\delta(k_1^2-A)\delta(k_2^2-m_c^2)\nonumber\\
&=&\frac{\partial}{\partial A }\frac{(-2\pi i)^2}{2\pi i}\int_{(\sqrt{A}+m_c)^2}^{\infty}dt\frac{1}{t-p^2}\frac{\pi}{2}\frac{\sqrt{\lambda(t,A,m_c^2)}}{t}\mid_{A\to m_c^2}\nonumber\\
&=&\frac{(-2\pi i)^2}{2\pi i}\int_{4m_c^2}^{\infty}dt\frac{1}{t-p^2}\frac{\pi}{2} \frac{-1}{\sqrt{t(t-4m_c^2)}}\, ,
\end{eqnarray}
divergence at the end-point $t=4m_c^2$ appears.  Such terms are companied with the small $s$-quark mass $m_s$ and play a minor important role, we regularize the endpoint divergences by adding a small squared $s$-quark mass $4m_s^2$ empirically, just like in previous works \cite{WZG-ZJX-Zc-Decay,WZG-Y4660-Decay,WZG-X4140-decay,WZG-X4274-decay,WZG-Z4600-decay}. If we vary the $4m_s^2$ by adding a small uncertainty $4m_s^2\pm\delta$,  and we can obtain almost invariant numerical results.

Furthermore,   we smear  the dependencies of the parameters  $C_{J/\psi K}^T$, $C_{J/\psi K}^M$, $C_{\eta_c K^*}^T$, $C_{\eta_c K^*}^M$, $C_{D^* \bar{D}_s}^T$,     and $C_{D^* \bar{D}_s}^M$ on the Lorentz invariants $p^{\prime2}$, $p^2$, $q^2$, and take them  as free parameters, and search for the best values  to
delete  the contaminations from the high resonances and continuum states to acquire  stable QCD sum rules with variations of
the Borel parameters $T^2$.

\section{Numerical results and discussions}	
At the QCD side, we take  the standard values
$\langle
\bar{q}q \rangle=-(0.24\pm 0.01\, \rm{GeV})^3$, $\langle\bar{s}s\rangle=(0.8\pm0.1)\langle\bar{q}q\rangle$,
$\langle\bar{q}g_s\sigma G q \rangle=m_0^2\langle \bar{q}q \rangle$,
$\langle\bar{s}g_s\sigma G s \rangle=m_0^2\langle \bar{s}s \rangle$,
$m_0^2=(0.8 \pm 0.1)\,\rm{GeV}^2$     at the   energy scale  $\mu=1\, \rm{GeV}$
\cite{SVZ79,Reinders85,Colangelo-Review},  and take the $\overline{MS}$ quark masses $m_{c}(m_c)=(1.275\pm0.025)\,\rm{GeV}$ and $m_s(\mu=2\,\rm{GeV})=(0.095\pm0.005)\,\rm{GeV}$ from the Particle Data Group \cite{PDG}. We set $m_u=m_d=0$ and take account of
the energy-scale dependence of  the input parameters,
\begin{eqnarray}
\langle\bar{q}q \rangle(\mu)&=&\langle\bar{q}q \rangle({\rm 1GeV})\left[\frac{\alpha_{s}({\rm 1GeV})}{\alpha_{s}(\mu)}\right]^{\frac{12}{33-2n_f}}\, , \nonumber\\
\langle\bar{s}s \rangle(\mu)&=&\langle\bar{s}s \rangle({\rm 1GeV})\left[\frac{\alpha_{s}({\rm 1GeV})}{\alpha_{s}(\mu)}\right]^{\frac{12}{33-2n_f}}\, , \nonumber\\
 \langle\bar{q}g_s \sigma Gq \rangle(\mu)&=&\langle\bar{q}g_s \sigma Gq \rangle({\rm 1GeV})\left[\frac{\alpha_{s}({\rm 1GeV})}{\alpha_{s}(\mu)}\right]^{\frac{2}{33-2n_f}}\, , \nonumber\\
  \langle\bar{s}g_s \sigma Gs \rangle(\mu)&=&\langle\bar{s}g_s \sigma Gs \rangle({\rm 1GeV})\left[\frac{\alpha_{s}({\rm 1GeV})}{\alpha_{s}(\mu)}\right]^{\frac{2}{33-2n_f}}\, , \nonumber\\
 m_c(\mu)&=&m_c(m_c)\left[\frac{\alpha_{s}(\mu)}{\alpha_{s}(m_c)}\right]^{\frac{12}{33-2n_f}} \, ,\nonumber\\
 m_s(\mu)&=&m_s({\rm 2GeV})\left[\frac{\alpha_{s}(\mu)}{\alpha_{s}({\rm 2GeV})}\right]^{\frac{12}{33-2n_f}} \, ,\nonumber\\
\alpha_s(\mu)&=&\frac{1}{b_0t}\left[1-\frac{b_1}{b_0^2}\frac{\log t}{t} +\frac{b_1^2(\log^2{t}-\log{t}-1)+b_0b_2}{b_0^4t^2}\right]\, ,
\end{eqnarray}
 from the renormalization group equation,   $t=\log \frac{\mu^2}{\Lambda_{QCD}^2}$, $b_0=\frac{33-2n_f}{12\pi}$, $b_1=\frac{153-19n_f}{24\pi^2}$, $b_2=\frac{2857-\frac{5033}{9}n_f+\frac{325}{27}n_f^2}{128\pi^3}$,  $\Lambda_{QCD}=210\,\rm{MeV}$, $292\,\rm{MeV}$  and  $332\,\rm{MeV}$ for the flavors  $n_f=5$, $4$ and $3$, respectively  \cite{PDG,Narison-mix},   and evolve all the input parameters to the typical  energy scale  $\mu=m_c(m_c)\approx 1.3\,\rm{GeV}$ to extract hadronic coupling constants \cite{WZG-Zcs3985-tetra,WZG-Zcs3985-mole,WZG-Hidden-charm-PRD}.

At the hadron side, we take the parameters  as $M_{K}=0.4937\,\rm{GeV}$,  $M_{K^*}=0.8917\,\rm{GeV}$,
$M_{J/\psi}=3.0969\,\rm{GeV}$, $M_{\eta_c}=2.9834\,\rm{GeV}$, $M_{D^*}=2.007\,\rm{GeV}$ and $D_{s}=1.969\,\rm{GeV}$ \cite{PDG},  $f_{K}=0.156\,\rm{GeV}$ \cite{PDG}, $f_{K^*}=0.220\,\rm{GeV}$, $\sqrt{u^0_{K}}=1.0\,\rm{GeV}$, $\sqrt{u^0_{K^*}}=1.3\,\rm{GeV}$ \cite{PBall-decay-Kv},
 $f_{D_s}=240\,\rm{MeV}$, $\sqrt{u^0_{D_s}}=2.6\,\rm{GeV}$,  $f_{D^*}=263\,\rm{MeV}$, $\sqrt{s^0_{D^*}}=2.5\,\rm{GeV}$  \cite{WZG-EPJC-decay},
$f_{J/\psi}=0.418 \,\rm{GeV}$, $f_{\eta_c}=0.387 \,\rm{GeV}$  \cite{Becirevic},  $\sqrt{s^0_{J/\psi}}=3.6\,\rm{GeV}$, $\sqrt{s^0_{\eta_c}}=3.5\,\rm{GeV}$ \cite{PDG}, $M_{Z}=3.99\,\rm{GeV}$,   $\lambda_{Z}^T=2.85\times 10^{-2}\,\rm{GeV}^5$ \cite{WZG-Zcs3985-tetra}, $\lambda_{Z}^M=1.96\times 10^{-2}\,\rm{GeV}^5$ \cite{WZG-Zcs3985-mole}, $\frac{f_{K}M^2_{K}}{m_u+m_s}=-\frac{\langle \bar{q}q\rangle+\langle \bar{s}s\rangle}{f_{K}(1-\delta_K)}$ from the Gell-Mann-Oakes-Renner relation, $\delta_K=0.50$ \cite{GMOR-fK}.

\begin{figure}
 \centering
 \includegraphics[totalheight=5cm,width=7cm]{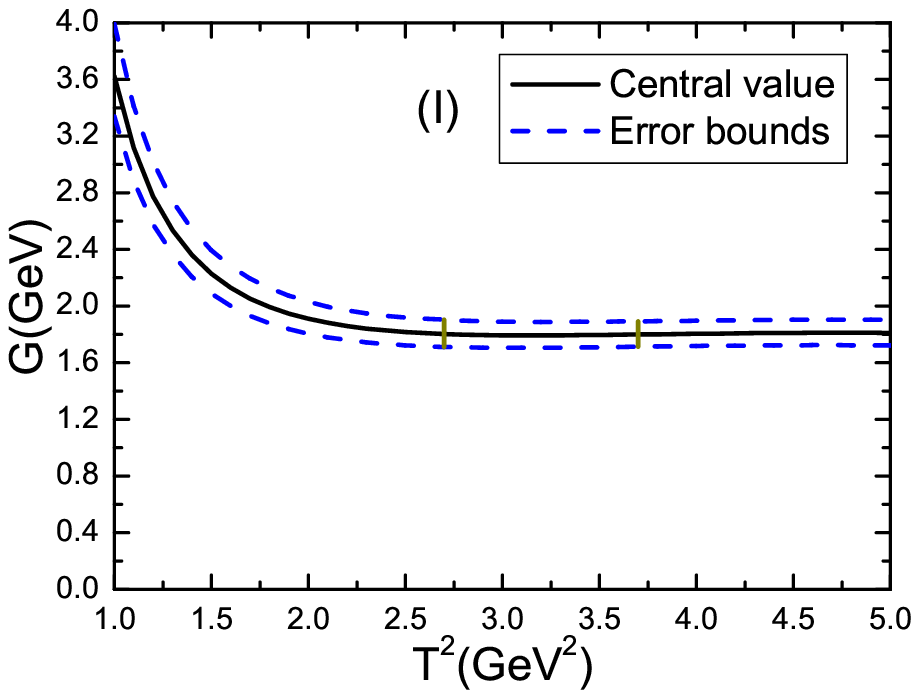}
 \includegraphics[totalheight=5cm,width=7cm]{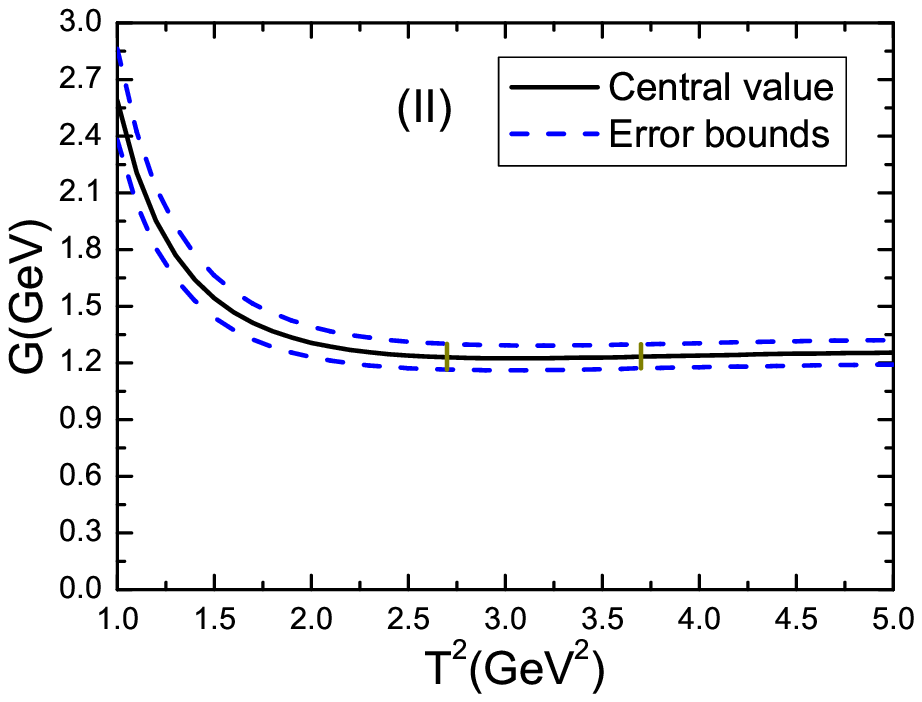}
  \caption{ The  hadronic coupling  constants $G^T_{ZJ/\psi K}$ (I) and $G^M_{ZJ/\psi K}$ (II)  with variations of the Borel parameter $T^2$, where the regions between the two vertical lines are the Borel windows.  }\label{ZcsPsiK-T-M}
\end{figure}

In calculations, we fit the unknown parameters to be   $C^T_{J/\psi K}=0.00045+0.00038\times T^2\,\rm{GeV}^8$,
 $C_{\eta_c K^*}^T=-0.00184-0.00068\times T^2\,\rm{GeV}^8$,  $C^T_{D^* \bar{D}_s}=0+0\times T^2\,\rm{GeV}^8$,
  $C^M_{J/\psi K}=0.0004+0.00017\times T^2\,\rm{GeV}^8$,
 $C^M_{\eta_c K^*}=0.00094+0.00033\times T^2\,\rm{GeV}^8$,
 and $C^M_{D^* \bar{D}_s}=-0.023-0.00405\times T^2\,\rm{GeV}^8$ to acquire the flat Borel platforms $T^2_{max}-T^2_{min}=1\,\rm{GeV}^2$, where the max and min represent the maximum and minimum values, respectively, see Fig.\ref{ZcsPsiK-T-M} for example.
In the  picture  of tetraquark  states,  the Borel parameters are $T^2_{J/\psi K}=(2.7-3.7)\,\rm{GeV}^2$, $T^2_{\eta_c K^*}=(2.5-3.5)\,\rm{GeV}^2$ and $T^2_{D^*\bar{D}_s}=(3.1-4.1)\,\rm{GeV}^2$ for the hadronic coupling constants $G_{ZJ/\psi K}^T$,
$G^T_{Z\eta_cK^*}$ and $G^T_{ZD^*\bar{D}_s}$, respectively, while in the  picture  of molecular  states, the Borel parameters $T^2_{J/\psi K}=(2.7-3.7)\,\rm{GeV}^2$, $T^2_{\eta_c K^*}=(2.3-3.3)\,\rm{GeV}^2$ and $T^2_{D^* \bar{D}_s}=(2.3-3.3)\,\rm{GeV}^2$ for the hadronic coupling constants $G_{ZJ/\psi K}^M$, $G^M_{Z\eta_cK^*}$ and $G^M_{ZD^*\bar{D}_s}$, respectively, where we add the subscripts $J/\psi K$, $\eta_c K^*$ and $D^*\bar{D}_s$ to denote  the corresponding channels.
The uncertainties originate from the uncertainties of the input parameters at the QCD side, in general, can be absorbed into the decay constants and coupling constants together, for example, the $f_{J/\psi}$, $f_{K}$, $\lambda_Z^T$ and $G^T_{ZJ/\psi K}$ in the QCD sum rules in Eq.\eqref{JpsiK-Z-T}.

If we use the symbol   $\xi$ to stand for the input parameters at the QCD side,  the uncertainties   $\xi \to \xi +\delta \xi$ lead to the uncertainties $f_{J/\psi}f_{K}\lambda^T_{Z}G^T_{ZJ/\psi K} \to f_{J/\psi}f_{K}\lambda^T_{Z}G^T_{ZJ/\psi K}+\delta\,f_{J/\psi}f_{K}\lambda^T_{Z}G^T_{ZJ/\psi K}$, $C^T_{J/\psi K} \to C^T_{J/\psi K}+\delta C^T_{J/\psi K}$,
where
\begin{eqnarray}\label{Uncertainty-4}
\delta\,f_{J/\psi}f_{K}\lambda^T_{Z}G^T_{ZJ/\psi K} &=&f_{J/\psi}f_{K}\lambda^T_{Z}G^T_{ZJ/\psi K}\left( \frac{\delta f_{J/\psi}}{f_{J/\psi}} +\frac{\delta f_{K}}{f_{K}}+\frac{\delta \lambda^T_{Z}}{\lambda^T_{Z}}+\frac{\delta G^T_{ZJ/\psi K}}{G^T_{ZJ/\psi K}}\right)\, ,
\end{eqnarray}
we can set $\frac{\delta f_{J/\psi}}{f_{J/\psi}} =\frac{\delta f_{K}}{f_{K}}=\frac{\delta \lambda^T_{Z}}{\lambda^T_{Z}}=\frac{\delta G^T_{ZJ/\psi K}}{G^T_{ZJ/\psi K}}
$ approximately, then
\begin{eqnarray}\label{Uncertainty}
\delta\,f_{J/\psi}f_{K}\lambda^T_{Z}G^T_{ZJ/\psi K} &=&f_{J/\psi}f_{K}\lambda^T_{Z}G^T_{ZJ/\psi K} \frac{4\delta G^T_{ZJ/\psi K}}{G^T_{ZJ/\psi K}} \, ,
\end{eqnarray}
to avoid overestimating the uncertainty of the hadronic coupling constant. In calculations, we estimate the uncertainties according to Eq.\eqref{Uncertainty} analogously and neglect the uncertainties of the unknown parameters $C^T_{J/\psi K}$,  $C_{\eta_c K^*}^T$,  $C^T_{D^* \bar{D}_s}$,  $C^M_{J/\psi K}$, $C^M_{\eta_c K^*}$,
 and $C^M_{D^* \bar{D}_s}$, except in the case of the $\delta m_c(m_c)$ for the $C^M_{D^* \bar{D}_s}$, where we have to take account of the uncertainty to acquire flat Borel platform.

Now let us obtain the values of the hadronic coupling constants routinely,
\begin{eqnarray}
G^T_{ZJ/\psi K} &=&1.79\pm0.10\,\rm{GeV}\, , \nonumber\\
|G^T_{Z\eta_cK^*}|&=&2.97\pm 0.22\,\rm{GeV}\, , \nonumber\\
|G^T_{ZD^*\bar{D}_s}|&=&0.71\pm0.04\,\rm{GeV}\, , \nonumber\\
G^M_{ZJ/\psi K} &=&1.23\pm0.07\,\rm{GeV}\, , \nonumber\\
G^M_{Z\eta_c K^*}&=&2.20\pm 0.16\,\rm{GeV}\, , \nonumber\\
|G^M_{ZD^*\bar{D}_s}|&=&9.60\pm1.27\,\rm{GeV}\, .
\end{eqnarray}

Then we choose the masses  $M_{K^-}=0.4937\,\rm{GeV}$,  $M_{K^{*+}}=0.8917\,\rm{GeV}$,
$M_{J/\psi}=3.0969\,\rm{GeV}$, $M_{\eta_c}=2.9834\,\rm{GeV}$,
$M_{D^{*0}}=2.0069 \,\rm{GeV}$, $M_{D^{-}_s}=1.9690\,\rm{GeV}$,  $M_{D^0}=1.8648 \,\rm{GeV}$, $M_{D^{*-}_s}=2.1122\,\rm{GeV}$ \cite{PDG},
  $M_{Z_c}=3.99\,\rm{GeV}$ \cite{WZG-Zcs3985-tetra,WZG-Zcs3985-mole}, and obtain the partial decay widths,
\begin{eqnarray}
\Gamma(Z_{cs}^T\to J/\psi K^-)&=&5.36\pm 0.60\,\rm{MeV}   \, ,\nonumber\\
\Gamma(Z_{cs}^T\to\eta_c K^{*-})&=&9.54\pm 1.41\,\rm{MeV}  \, , \nonumber\\
\Gamma(Z_{cs}^T\to D^{*0} \bar{D}^{-}_s)&=&0.21\pm 0.02\,\rm{MeV}   \, ,\nonumber\\
\Gamma(Z_{cs}^T\to D^0 \bar{D}^{*-}_s)&=&0.20\pm 0.02\,\rm{MeV}   \, ,
\end{eqnarray}
\begin{eqnarray}
\Gamma(Z_{cs}^M\to J/\psi K^-)&=&2.53\pm 0.29\,\rm{MeV}   \, ,\nonumber\\
\Gamma(Z_{cs}^M\to\eta_c K^{*-})&=&5.23\pm 0.76 \,\rm{MeV}  \, , \nonumber\\
\Gamma(Z_{cs}^M\to D^{*0} \bar{D}^{-}_s)&=&38.69\pm 10.24\,\rm{MeV}   \, ,\nonumber\\
\Gamma(Z_{cs}^M\to D^0 \bar{D}^{*-}_s)&=&37.06\pm 9.81\,\rm{MeV}   \, ,
\end{eqnarray}
and the total widths,
\begin{eqnarray}
\Gamma_{Z_{cs}}^T &=&15.31\pm 2.06\,\rm{MeV}\, ,\nonumber\\
\Gamma_{Z_{cs}}^M &=& 83.51\pm21.09\,\rm{MeV}\, ,
\end{eqnarray}
which are consistent with the experimental data $13.8^{+8.1}_{-5.2}\pm4.9\,\rm{MeV}$ from the BESIII collaboration \cite{BES3985} and
$131 \pm 15 \pm 26\,\rm{MeV}$ from the LHCb collaboration \cite{LHCb-Zcs4000}, respectively.
 There are two $Z_{cs}$ states with analogous masses, just like in the case of the $Z_c^{\pm}$ states, there exist a tetraquark candidate $Z_c(3900)$ and a molecule candidate $Z_c(3885)$ \cite{WZG-ZJX-Zc-Decay}.

 The physical $Z_{cs}$ states maybe have both the diquark-antidiquark type and color-singlet-color-singlet type tetraquark Fock components, it is better to choose the mixing currents or more physical currents to interpolate them,
 \begin{eqnarray}
 J_\nu(x)&=&J^T_\nu(x)\,\cos\theta+J^M_\nu(x)\,\sin\theta\, ,
 \end{eqnarray}
with the mixing angle $\theta$. As both the currents $J^T_\nu(x)$ and $J^M_\nu(x)$ can lead to a mass about $3.99\pm0.09\,\rm{GeV}$ in the case of choosing the same constraints \cite{WZG-Zcs3985-tetra,WZG-Zcs3985-mole}. The additional parameter $\theta$ cannot affect the predicted mass remarkably, but can affect the predicted width remarkably. We can estimate the hadronic coupling constants $G$ with the simple replacement,
\begin{eqnarray}
G&\to&G^T\,\cos\theta+G^M\,\sin\theta\, ,
\end{eqnarray}
and compare the predicted total widths and  branching fractions to the precise experimental data in the future to estimate  the mixing angle $\theta$, the branching fractions have not been measured yet. The significance of the $Z^{-}_{cs}(3985)$ is about $5.3 \, \sigma$   \cite{BES3985}, while the significance of the $Z^{+}_{cs}(4000)$ is about $15\,\sigma$ \cite{LHCb-Zcs4000}, such large significances only indicate that there maybe exist those  resonance structures indeed, more experimental data are still needed  even to distinguish (and confirm) the $Z_{cs}(3985)$ and $Z_{cs}(4000)$ unambiguously. In the present work, at least, we can obtain the conclusion tentatively that larger color-singlet-color-singlet component in the $Z_{cs}$ state  leads to larger decay width, the $Z_{cs}(3985)$ maybe have large diquark-antidiquark type Fock component, while the $Z_{cs}(4000)$ maybe have large color-singlet-color-singlet type Fock component.

\section{Conclusion}
In this work, we explore the hadronic coupling constants $G_{ZJ/\psi K}$, $G_{Z\eta_c K^*}$, $G_{ZD^* \bar{D}_s}$ of the exotic states $Z_{cs}(3985/4000)$ both in the pictures of the tetraquark states and molecular states with the tentative assignments $J^{PC}=1^{+-}$ based on the rigorous quark-hadron duality.   We write down   the three-point correlation functions, and accomplish the operator product expansion up to the vacuum condensates of dimension-5, and neglect the tiny gluon condensate contributions, just like in our previous works, and obtain the spectral densities through dispersion relation,    then we acquire the rigorous quark-hadron duality below the continuum thresholds $s_0$ and $u_0$ by accomplishing  the integral over the variable $ds^\prime$ firstly, and obtain six  QCD sum rules for the hadronic coupling constants.      We investigate  the two-body strong decays
of the axialvector tetraquark state and molecular state, respectively,  and obtain the total widths
$\Gamma_{Z_{cs}}^T =15.31\pm 2.06\,\rm{MeV}$ and $\Gamma_{Z_{cs}}^M=83.51\pm21.09\,\rm{MeV}$,
which are consistent with the experimental data $13.8^{+8.1}_{-5.2}\pm4.9\,\rm{MeV}$ from the BESIII collaboration  and
$131 \pm 15 \pm 26\,\rm{MeV}$ from the LHCb collaboration, respectively. The present calculations support assigning the $Z_{cs}(3985)$ and $Z_{cs}(4000)$ to be the hidden-charm tetraquark state and molecular state with the $J^{PC}=1^{+-}$, respectively. Or at least, the $Z_{cs}(3985)$ maybe have large diquark-antidiquark type Fock component, while the $Z_{cs}(4000)$ maybe have large color-singlet-color-singlet type Fock component.

\section*{Acknowledgements}
This  work is supported by National Natural Science Foundation, Grant Number  12175068.

\end{document}